\documentclass[aps,prl,twocolumn,superscriptaddress,groupedaddress]{revtex4}  % for review and submission
\usepackage{graphicx}  % needed for figures
\usepackage{dcolumn}   % needed for some tables
\usepackage{bm}        % for math
\usepackage{amssymb}   % for math

\begin{document}

% The following information is for internal review, please remove them for submission
%\widetext
%\leftline{Version xx as of \today}
%\leftline{Primary authors: Joe E. Physics}
%\leftline{To be submitted to (PRL, PRD-RC, PRD, PLB; choose one.)}
%\leftline{Comment to {\tt d0-run2eb-nnn@fnal.gov} by xxx, yyy}
%\centerline{\em D\O\ INTERNAL DOCUMENT -- NOT FOR PUBLIC DISTRIBUTION}

% the following line is for submission, including submission to the arXiv!!
%\hspace{5.2in} \mbox{Fermilab-Pub-04/xxx-E}

\title{(2+1)-dimensional interacting model of two massless spin-2 fields as a bi-gravity model}
%\input author_list.tex       % D0 authors (remove the first 3 lines
                             % of this file prior to submission, they
                             % contain a time stamp for the authorlist)
                             % (includes institutions and visitors)
\author { \small{ \bf S. Hoseinzadeh } { \small 
and}
\small{ \bf A. Rezaei-Aghdam } \\
{\small{\em Department of Physics, Faculty of Science, Azarbaijan Shahid Madani University, }}\\
{\small{\em 53714-161, Tabriz, Iran }}}

\date{\today}

\begin{abstract}
We propose a new group-theoretical (Chern-Simons) formulation for the bi-metric theory of gravity in (2+1)-dimensional spacetime which describe two interacting massless spin-2 fields. Our model has been formulated in terms of two dreibeins 
rather than two metrics.
We obtain our Chern-Simons gravity model by gauging {\it mixed AdS-AdS Lie algebra} and show that it has a two dimensional conformal field theory (CFT) at the boundary of the anti de Sitter (AdS) solution. We show that the central charge of the dual CFT is proportional to the mass of the AdS solution. We also study cosmological implications of our massless bi-gravity model.
\end{abstract}

%\pacs{04.60.Kz, 04.70.Bw, 11.15.-q}
\maketitle

%%%%%%%%%%%%%%%%%%%%%%%%%%%%%%%%%%%%% 
\section{I. Introduction}

Einstein's general relativity is a field theory which describes interactions of a single massless spin-2 particle (graviton). Because many of the proposal theories for quantum gravity such as M-theory and (super)string theory have a massless graviton in their spectrum, obtaining a quantum theory of massless interacting spin-2 fields may be a useful way for solving the problem of finding a consistent quantum theory of gravity. It has been proved that does not exist any consistent theory (with at most two derivatives of fields) involving interactions of many massless spin-2 fields in spacetime dimensions $d\!>\!3$, because an unphysical scalar mode of negative energy (ghost) appears in such theories \cite{1Boulanger}. Although, in (2+1)-dimensional spacetime a consistent interacting theory of many massless spin-2 fields has been constructed in \cite{2Boulanger}, but physical consequences of such a (2+1)-dimensional interacting model of massless spin-2 fields has not been studied in detail. 
Theories which describe massless spin-2 fields in (2+1)-dimensional spacetime, have no local degrees of freedom (DOF), hence the Chern-Simons theory is a suitable candidate to construct a (2+1)-dimensional interacting theory of massless spin-2 fields.
(2+1)-dimensional Chern-Simons theory (with any gauge group) is a topological model and has no local DOF.
Part of interests in (2+1)-dimensional gravity come from simplifying features of the quantum theory arising from the lower dimension.
In particular, (2+1)-dimensional (Chern-Simons) gravity models (with negative cosmological constant) could have $AdS_{3}/CFT_{2}$ correspondence, in which a (2+1)-dimensional quantum gravity in an asymptotically AdS spacetime is equivalent to a two dimensional CFT, defined on the asymptotic boundary of AdS spacetime.
In (2+1)-dimensional spacetime, various Chern-Simons gravity models such as general relativity with zero, negative or positive cosmological constants, conformal gravity, Maxwell and AdS-Lorentz gravities have been constructed by gauging their appropriate gauge groups \cite{1E.Witten,1S.Carlip,2S.Carlip,3S.Carlip,2E.Witten,Horne,J.Diaz,P.Salgado,1S.Hoseinzadeh}. 
Moreover, it has been shown that some three dimensional massive gravity models such as
Fierz-Pauli theory \cite{Fierz}, topologically massive gravity \cite{1S.Deser,2S.Deser,4S.Carlip,1M.Blagojevic}, new massive gravity  \cite{1E.A.Bergshoeff,2E.A.Bergshoeff,2M.Blagojevic,O.Hohm},
zwei-dreibein gravity (ZDG) \cite{3E.A.Bergshoeff}, and general zwei-dreibein gravity (GZDG) \cite{5E.A.Bergshoeff}, have Chern-Simons-like formulations \cite{W.Merbis}, and in contrary to the Chern-Simons theories have local DOF and propagating massive spin-2 modes. Both ZDG and GZDG describe two interacting massive spin-2 modes. 

On the other hand, to understand the recent observational data in cosmology which indicate that the expansion of the universe is accelerating, it is common to formulate new theories of the gravitational interaction, and study their cosmological implications.
A natural possibility for extending known classical field theories is to include additional spin-2 fields and interactions.
Including spin-2 interactions modifies general relativity at large distances. Bi-metric theory is therefore a candidate to explain the accelerated expansion of the universe.
The interactions could change some of the dynamics of the gravitational theory which could be relevant for cosmological applications. This is one of the motivations of massive and bi-metric theories of gravity, which changes the long-distance behaviour of the gravitational field, making them candidate theories of dark matter and energy.

In this paper, we propose a new group-theoretical (Chern-Simons) formulation for the bi-metric theory of gravity in (2+1)-dimensional spacetime such that it describes two interacting massless spin-2 particles.
Our “bi-metric” model, similar to the ZDG model, has been formulated in terms of two dreibeins rather than two metric tensors. But, unlike the ZDG which is a massive gravity model, our model can be viewed as a massless “zwei-dreibein gravity” theory.

In section two, we introduce a new Lie algebra with two anti de Sitter subalgebras in 2+1 spacetime dimensions.
Then, we construct a Chern-Simons gravity model using the obtained Lie algebra as a gauge algebra. Our model contains a pair of dreibeins together with their corresponding spin connections. Our ghost-free model which has a suitable free field limit, has no local degrees of freedom and describes two interacting massless spin-2 fields.
We obtain an AdS solution to the equations of motion involving a pair of AdS spacetime metrics, and calculate the mass of the solution.
In section three, we show that the model has a dual CFT at the boundary of the AdS spacetime solution, whose central charge is proportional to the mass of the AdS solution. We show that this result is valid also in the ZDG model.
In section four, we discuss about cosmological implications of the model, and show that it admits some homogeneous and isotropic Friedmann-Robertson-Walker solutions.
Section five is devoted to the concluding remarks.

%%%%%%%%%%%%%%%%%%%%%%%%%%%%%%%%%%%%%%%
\section{II. (2+1)-dimensional Chern-Simons bi-gravity model}

In this section, we introduce a new Lie algebra in (2+1)-dimensional spacetime, which has two anti de Sitter subalgebras.
In (2+1)-dimensional spacetime, the commutation relations for the 12-dimensional \textit{mixed AdS-AdS Lie algebra} has the following form:
\begin{eqnarray} 
[J_{a},P_{b}] \!=\! \epsilon_{abc} P^{c},~~~
[J_{a},J_{b}] \!=\! \ell^{2}[P_{a},P_{b}] \!=\!\epsilon_{abc} J^{c}, ~~~~~ 
\nonumber
\\ \nonumber
[\tilde{J}_{a},\tilde{P}_{b}] \!=\! \epsilon_{abc} \tilde{P}^{c},~~~ 
[\tilde{J}_{a},\tilde{J}_{b}] \!=\! \tilde{\ell}^{2}[\tilde{P}_{a},\tilde{P}_{b}] \!=\!\epsilon_{abc} \tilde{J}^{c},~~~~~
\\ \nonumber
[P_{a},\tilde{J}_{b}] \!=\!\ell^{-1}[J_{a},\tilde{J}_{b}] \!=\! (2\ell)^{-1} \epsilon_{abc} (J^{c}+\ell P^{c}),~~~~~~~~~
\\ \label{mixed AdS-AdS Lie algebra}
~~[J_{\!a},\!\tilde{P}_{b}] \!\!=\!\ell[P_{\!\!a},\!\tilde{P}_{b}] \!\!=(2\tilde{\ell})^{\!-\!1}\!\epsilon_{abc} \!\Big(\!(\! J^{c}\!\!+\!\ell P^{c})\!-\!2(\!\tilde{J}^{c}\!\!-\!\tilde{\ell} \tilde{P}^{c}) \!\Big)\!,
\end{eqnarray}
where $\ell^{-2}\!\!=\!-\Lambda$ and $\tilde{\ell}^{-2}\!=\!\!-\tilde{\Lambda}$ are constants, $\epsilon_{012}=-1$, and $J_{a},\tilde{J}_{a}$  and $P_{a},\tilde{P}_{a}$ ($a=0,1,2$) are the Lorentz and translation generators, respectively. The algebra indices $a,b,c$ can be rised and lowered by (2+1)-dimensional Minkowski metric $\eta_{ab}=diag(-1,1,1)$.
In what follows, we use the mixed AdS-AdS Lie algebra (\ref{mixed AdS-AdS Lie algebra}) to construct a (2+1)-dimensional Chern-Simons gravity model.
The Lie algebra (\ref{mixed AdS-AdS Lie algebra}) admits the following bilinear quadratic form \cite{C.Nappi}:
\begin{eqnarray} \label{adinvariant metric} 
\langle J_{\!a},\! P_{b}\rangle \!\!=\!\!(\!2\beta\!-\!\gamma\!)\ell^{-\!1} \eta_{ab},~~
\langle P_{\!a},\! P_{b}\rangle \!\!=\!\!\ell^{-\!2}\! \langle J_{\!a},\! J_{\!b}\rangle 
\!\!=\!\!\gamma \ell^{-\!2} \eta_{ab},~ \nonumber
\\
\langle \tilde{J}_{a},\!\tilde{P}_{b}\rangle \!\!=\!\!(\!\alpha\!-\!\beta\!)\tilde{\ell}^{-\!1} \eta_{ab},~~~
\langle \tilde{P}_{a},\!\tilde{P}_{b}\rangle \!\!=\!\!\tilde{\ell}^{-\!2}\!\langle \tilde{J}_{a},\!\tilde{J}_{b}\rangle \!\!=\!\!\alpha\tilde{\ell}^{-\!2} \eta_{ab}, \nonumber
\\
\langle J_{a},\!\tilde{P}_{b}\rangle \!\!=\!\!\langle P_{\!a},\!\tilde{P}_{b}\rangle \!\!=\! 0,~~~~~
\langle P_{\!a},\!\tilde{J}_{b}\rangle \!\!=\!\ell^{-\!1}\!\langle J_{a},\!\tilde{J}_{b}\rangle 
\!\!=\!\!\beta\ell^{-\!1}\! \eta_{ab},~
\end{eqnarray}
where $\alpha$, $\beta$ and $\gamma$ are arbitrary constants. The ad-invariant metric (\ref{adinvariant metric}) is non-degenerate for $\beta\neq0$, $\beta\neq\alpha$, $\beta\neq\gamma$.
The mixed AdS-AdS Lie algebra valued Murer-Cartan one-form gauge field $h=h_{\mu}~dx^{\mu}$ can be written as:
\begin{eqnarray} \label{Murer-Cartan one-form}
h_{\mu}=h_{\mu}^{~B}X_{B}=e_{\mu}^{~a}P_{a}+\omega_{\mu}^{~a}J_{a} +\tilde{e}_{\mu}^{~a}\tilde{P}_{a}+\tilde{\omega}_{\mu}^{~a}\tilde{J}_{a},
\end{eqnarray}
where the Greek indices $\mu=0,1,2$ are the spacetime indices, $e_{\mu}^{~a}, \omega_{\mu}^{~a}$ are the dreibein and spin
connection corresponding to the generators $P_{a},J_{a}$, and $\tilde{e}_{\mu}^{~a} , \tilde{\omega}_{\mu}^{~a}$ are dreibein and spin
connection corresponding to the generators $\tilde{P}_{a},\tilde{J}_{a}$, respectively.
The curvature two-form tensor corresponding to the algebra (\ref{mixed AdS-AdS Lie algebra}) has the following form:
\begin{eqnarray} \label{curvature two-form}
\mathcal{R}_{\mu\nu}=\partial_{[\mu}h_{\nu]}+[h_{\mu},h_{\nu}] =
\mathcal{R}_{\mu\nu}^{~A} X_{A}~~~~~~~~~~~~~~~~   \nonumber
\\
=T_{\mu\nu}^{~~a}~P_{a}+R_{\mu\nu}^{~~a}~J_{a} +\tilde{T}_{\mu\nu}^{~~a}~\tilde{P}_{a}+\tilde{R}_{\mu\nu}^{~~a}~\tilde{J}_{a},
\end{eqnarray}
where the torsion tensors $T_{\mu\nu}^{~~a}$ and $\tilde{T}_{\mu\nu}^{~~a}$, and the Riemannian curvatures $R_{\mu\nu}^{~~a}$ and $\tilde{R}_{\mu\nu}^{~~a}$ are obtained as follows:
\begin{eqnarray}  \label{Torsion-Curvatures}
T_{\mu\nu}^{~~c}=D_{\mu} e_{\nu}^{~c} 
+\frac{\ell}{2}\epsilon_{ab}^{~~c}\Big(\frac{1}{\ell\tilde{\ell}} e_{[\mu}^{~a} \tilde{e}_{\nu]}^{~b}   
+\omega_{[\mu}^{~a}  \tilde{\omega}_{\nu]}^{~b}~~~~~~~~~~~~~~~~~
\nonumber
\\ 
+\frac{1}{\ell} e_{[\mu}^{~a} \tilde{\omega}_{\nu]}^{~b}
+\frac{1}{\tilde{\ell}} \omega_{[\mu}^{~a} \tilde{e}_{\nu]}^{~b}\Big),~~~~~~~~~~~~~~~~~~~~~~~~~~~~~~~~~   \nonumber
\\ 
R_{\mu\nu}^{~~c}=D_{\mu} \omega_{\nu}^{~c} +\frac{1}{2} \epsilon_{ab}^{~~c}\Big(  \frac{2}{\ell^{2}}  e_{\mu}^{~a} e_{\nu}^{\ b} 
+\frac{1}{\ell\tilde{\ell}} e_{[\mu}^{~a} ~\tilde{e}_{\nu]}^{~b}~~~~~~~~~~~~~  \nonumber
\\ 
+\frac{1}{\ell} e_{[\mu}^{~a} ~\tilde{\omega}_{\nu]}^{~b}
+\frac{1}{\tilde{\ell}} \omega_{[\mu}^{~a} ~\tilde{e}_{\nu]}^{~b}
+\omega_{[\mu}^{~a}~\tilde{\omega}_{\nu]}^{~b}\Big),~~~~~~~~~~~~~~~~~  \nonumber 
\\ 
\tilde{T}_{\mu\nu}^{~~c}=\tilde{D}_{\mu} \tilde{e}_{\nu}^{~c} 
+\epsilon_{ab}^{~~c}\Big(\frac{1}{\ell} e_{[\mu}^{~a}~\tilde{e}_{\nu]}^{~b} 
+\omega_{[\mu}^{~a}~\tilde{e}_{\nu]}^{~b}\Big),~~~~~~~~~~~~~~~ \nonumber
\\ 
\tilde{R}_{\mu\nu}^{~~c}=\tilde{D}_{\mu} \tilde{\omega}_{\nu}^{~c} +\epsilon_{ab}^{~~c}\Big(\frac{1}{\tilde{\ell}^{2}} \tilde{e}_{\mu}^{~a} \tilde{e}_{\nu}^{\ b} 
\!-\! \frac{1}{\ell\tilde{\ell}} e_{[\mu}^{~a}  \tilde{e}_{\nu]}^{~b} 
\!-\!\frac{1}{\tilde{\ell}} \omega_{[\mu}^{~a}  \tilde{e}_{\nu]}^{~b}\Big),~
\end{eqnarray}
where the covariant derivatives with respect to the spin connections $\omega_{\mu}^{~a}$ and $\tilde{\omega}_{\mu}^{~a}$ have the following forms:
\begin{eqnarray}
D_{\nu} \omega_{\rho c}=\partial_{[\nu} \omega_{\rho]c}+
\epsilon_{abc} \omega_{\nu}^{\ a} \omega_{\rho}^{\ b}, \nonumber
\\
D_{\nu} e_{\rho c}=\partial_{[\nu} e_{\rho]c}+
\epsilon_{abc} \omega_{[\nu}^{\ a} e_{\rho]}^{\ b},~  \nonumber
\\
\tilde{D}_{\nu} \tilde{\omega}_{\rho c}=\partial_{[\nu} \tilde{\omega}_{\rho]c}+
\epsilon_{abc} \tilde{\omega}_{\nu}^{\ a} \tilde{\omega}_{\rho}^{\ b},  \nonumber
\\
\tilde{D}_{\nu} \tilde{e}_{\rho c}=\partial_{[\nu} \tilde{e}_{\rho]c}+
\epsilon_{abc} \tilde{\omega}_{[\nu}^{\ a} \tilde{e}_{\rho]}^{\ b}.~
\end{eqnarray} 
We use the ad-invariant metric (\ref{adinvariant metric}), the one-form gauge field (\ref{Murer-Cartan one-form}), and the Chern-Simons action: 
$I_{cs}=\frac{1}{4\pi}\int_{M} \Big(\langle h \wedge dh \rangle +
\frac{1}{3}~\langle h \wedge [h \wedge h]\rangle \Big),$
to construct a (2+1)-dimensional gauge invariant Chern-Simons bi-metric gravity as follows:
\begin{eqnarray}\label{First model} 
I={I}^{e,\omega}(e,\omega)+{I}^{\tilde{e},\tilde{\omega}}(\tilde{e},\tilde{\omega})+I^{int}(e,\omega,\tilde{e},\tilde{\omega}),
\end{eqnarray}
where two first terms are
\begin{eqnarray}\nonumber
{I}^{e,\omega}\!\!\!=\!4G\frac{\gamma\!\!-\!\!2\beta}{\ell}\! I_{\!EC}\!(\!\omega,\!e,\!\Lambda\!)
\!+\!\gamma \!\!\!\int_{\!M}\!\!\! \frac{d^{3}\!x}{4\pi\!} \!\Big\{\! 2 \mathcal{L}_{\!L\!C\!S\!}(\!\omega\!)
\!+\!\! \frac{\!\epsilon^{\mu\nu\rho}\!}{2\ell^{2}} e_{\!\mu}^{c} D_{\!\nu} e_{\!\rho c}\!\Big\}\!, 
\nonumber
\\ \nonumber
I^{\tilde{e},\tilde{\omega}}\!\!\!=\!4G\frac{\beta\!\!-\!\alpha}{\tilde{\ell}}\! I_{\!EC}\!(\!\tilde{\omega},\!\tilde{e},\!\tilde{\Lambda}\!)
\!+\!\alpha\!\!\! \int_{\!M}\!\!\! \frac{d^{3}\!x}{4\pi\!} \!\Big\{\! 2 \mathcal{L}_{\!L\!C\!S\!}(\!\tilde{\omega}\!)
\!+\!\! \frac{\!\epsilon^{\mu\nu\rho}\!}{2\tilde{\ell}^{2}} \tilde{e}_{\!\mu}^{c} \tilde{D}_{\!\nu} \tilde{e}_{\!\rho c}\!\Big\}\!,
\end{eqnarray}
and the third term contains interactions of the fields $e_{\mu}^{~a}, \omega_{\mu}^{~a}$ with the fields $\tilde{e}_{\mu}^{~a} , \tilde{\omega}_{\mu}^{~a}$:
\begin{eqnarray}\nonumber
I^{int}\!\!=\! \beta\!\!\!\int_{\!M}\!\!\! \frac{d^{3}\!x}{4\pi} \epsilon^{\mu\nu\rho} \Big\{\!
\tilde{\omega}_{\mu}^{~c} D_{\nu} (\omega_{\rho c}\!+\!\ell^{-\!1} e_{\rho c})
\!+\!\ell^{-\!2} \epsilon_{abc} e_{\nu}^{~a}e_{\rho}^{~b}\tilde{\omega}_{\mu}^{~c}
\nonumber
\\ \nonumber
\!+\epsilon_{abc} (\tilde{\omega}_{\nu}^{~a}\tilde{\omega}_{\rho}^{~b}\!+\!\tilde{\ell}^{-\!2} \tilde{e}_{\nu}^{~a}\tilde{e}_{\rho}^{~b})(\omega_{\mu}^{~c}\!+\!\ell^{-\!1} e_{\mu}^{~c}) \!\Big\},
\end{eqnarray}
where $I_{EC}$ is the Einstein-Cartan Theory (EC) in 2+1 dimensions as follows:
\[
I_{EC}(\omega,e,\Lambda)\!=\!-\frac{1}{16\pi G}\!\!\int\!\! d^{3}x \epsilon^{\mu\nu\rho}
e_{\mu}^{~c}\Big(\! D_{\nu} \omega_{\rho c}
-\frac{\Lambda}{3}\epsilon_{abc} e_{\nu}^{~a}
e_{\rho}^{~b} \!\Big),
\]
and similarly for $I_{EC}(\tilde{\omega},\tilde{e},\tilde{\Lambda})$, and the Lorentz-Chern-Simons term $\mathcal{L}_{LCS}$, has the
following form:
\begin{eqnarray}\label{Lorentz-Chern-Simons term}
\mathcal{L}_{LCS}(\omega)=\frac{1}{4} \epsilon^{\mu\nu\rho}~\omega_{\mu}^{~c}
(\partial_{[\nu}~\omega_{\rho]c}+ \frac{2}{3}\epsilon_{abc}
~\omega_{\nu}^{\ a} \omega_{\rho}^{\ b}).
\end{eqnarray} 
The action (\ref{First model}) is a bi-metric theory of gravity (bi-gravity) which contains two spacetime metrics $g_{\mu\nu}=e_{\mu}^{~a} e_{\nu}^{~b} \eta_{ab}$ and $f_{\mu\nu}=\tilde{e}_{\mu}^{~a} \tilde{e}_{\nu}^{~b} \eta_{ab}$, correponding to two dreibeins $e_{\mu}^{~a}$ and $\tilde{e}_{\mu}^{~a}$, respectively.
The action (\ref{First model}) is a gauge invariant pure Chern-Simons action without any local degrees of freedom, and then it is a ghost-free model which describes two interacting massless spin-2 fields in 2+1 spacetime dimensions.
For $\beta\rightarrow 0,$\footnote{By letting $\beta\rightarrow 0$, the ad-invariant metric (\ref{adinvariant metric}) becomes degenerate, but this is just the free limit of the full action, without interaction terms.} one obtains free field limit of the action (\ref{First model}) as follows:
\begin{eqnarray}\label{free field limit}
I_{free}={I}^{e,\omega}(e,\omega,\beta=0)+{I}^{\tilde{e},\tilde{\omega}}(\tilde{e},\tilde{\omega},\beta=0),
\end{eqnarray}
which is also a ghost-free Chern-Simons action without local degrees of freedom, and describes two non-interacting massless spin-2 fields.
For $\alpha=\gamma=0,~\beta\neq0$ in (\ref{adinvariant metric}), one obtains another bi-metric Chern-Simons action, smaller than (\ref{First model}) as follows:
\begin{eqnarray}\label{second model}
I'\!=\! -4G\beta \Big(\frac{2}{\ell} I_{EC}(\omega,e,\Lambda)\! -\!\frac{1}{\tilde{\ell}} I_{EC}(\tilde{\omega},\tilde{e},\tilde{\Lambda})\Big)  \!+\! I^{int},~~
\end{eqnarray}
which is a ghost-free pure Chern-Simons action similar to (\ref{First model}), and has no local degrees of freedom. It describes two interacting massless spin-2 fields. In absence of the interaction terms, the free field limit of the action (\ref{second model}) is exactly the sum of two independent Einstein-Cartan actions:
\begin{eqnarray}\label{second free field limit}
I'_{free}=-4G\beta \Big(\frac{2}{\ell} I_{EC}(\omega,e,\Lambda)
-\frac{1}{\tilde{\ell}} I_{EC}(\tilde{\omega},\tilde{e},\tilde{\Lambda})\Big),~
\end{eqnarray}
which describes two non-interacting massless spin-2 fields \cite{1Boulanger}.
Although the model (\ref{First model}) is an interacting model of two massless spin-2 fields, but for a special values of the constants $\alpha,\beta,\gamma$ in the ad-invariant metric (\ref{adinvariant metric}), it can be 
changed to a non-interacting model describing two non-interacting massless spin-2 fields. In fact, for $\alpha=\gamma=\frac{\beta}{2}$, the following map:
\begin{eqnarray}\label{map}
e_{\mu}^{~a} \rightarrow \frac{\ell}{2} \Big(-\frac{2}{\ell} e_{\mu}^{~a} -\tilde{\omega}_{\mu}^{~a} +\frac{1}{\tilde{\ell}} \tilde{e}_{\mu}^{~a}\Big),~~~~  \nonumber
\\
\omega_{\mu}^{~a} \rightarrow \frac{1}{2} \Big(-\frac{2}{\ell} e_{\mu}^{~a} +\tilde{\omega}_{\mu}^{~a} -\frac{1}{\tilde{\ell}} \tilde{e}_{\mu}^{~a}\Big),~~~~  \nonumber
\\
\tilde{e}_{\mu}^{~a} \rightarrow \frac{\tilde{\ell}}{2} \Big(\omega_{\mu}^{~a}-\frac{1}{\ell} e_{\mu}^{~a} -\tilde{\omega}_{\mu}^{~a} -\frac{1}{\tilde{\ell}} \tilde{e}_{\mu}^{~a}\Big),  \nonumber
\\
\tilde{\omega}_{\mu}^{~a} \rightarrow \frac{1}{2} \Big(\omega_{\mu}^{~a}+\frac{3}{\ell} e_{\mu}^{~a} +\tilde{\omega}_{\mu}^{~a} +\frac{1}{\tilde{\ell}} \tilde{e}_{\mu}^{~a}\Big),
\end{eqnarray}
transforms the model (\ref{First model}) to the sum of two independent Einstein-Cartan actions (similar to (\ref{second free field limit})) \cite{1Boulanger}.
Assuming the following relations among fields and constants:
\begin{eqnarray}
e_{\mu}^{~a} \equiv e_{1\mu}^{~a},~~~~ \omega_{\mu}^{~a} \equiv \omega_{1\mu}^{~a},~~~~ \tilde{e}_{\mu}^{~a} \equiv e_{2\mu}^{~a},~~~~ \tilde{\omega}_{\mu}^{~a} \equiv \omega_{2\mu}^{~a},  \nonumber 
\\ 
\gamma=\frac{M_{p}(2\beta_{2}+\sigma\alpha_{2})}{\hat{m}\alpha_{2}\sqrt{\sigma\alpha_{1}}},~~~ 
\alpha=\frac{M_{p}}{\hat{m}\alpha_{2}}\Big(\frac{\beta_{2}}{\sqrt{\sigma\alpha_{1}}}-\sqrt{\alpha_{2}}\Big),  \nonumber
\\
\beta=\frac{M_{p}\beta_{2}}{\hat{m}\alpha_{2}}\sqrt{\frac{\sigma}{\alpha_{1}}},~~~
\Lambda=-\frac{\alpha_{1}\hat{m}^{2}}{\sigma},~~~ \tilde{\Lambda}=-\alpha_{2}\hat{m}^{2},~~~
\end{eqnarray}
the action (\ref{First model}) can be rewritten as follows:
\begin{eqnarray} 
I=\frac{1}{2\pi} I_{GZDG} +\frac{1}{4\pi} \!\!\int\!\! d^{3}x  \epsilon^{\mu\nu\rho} \Big\{   \beta \tilde{\omega}_{\mu}^{~c} D_{\nu} (\omega_{\rho c}\!+\!\frac{1}{\ell} e_{\rho c})~~~~ \nonumber
\\
 +\beta \epsilon_{abc} \Big(\! \tilde{\omega}_{\nu}^{~a}\tilde{\omega}_{\!\rho}^{~b}(\omega_{\!\mu}^{\!~c}\!+\!\frac{1}{\ell} e_{\!\mu}^{\!~c}) \!+\!\frac{1}{\ell^{2}} e_{\nu}^{~a}e_{\!\rho}^{\!~b}\tilde{\omega}_{\!\mu}^{\!~c} \!+\!\frac{1}{\tilde{\ell}^{2}} \tilde{e}_{\nu}^{~a}\tilde{e}_{\!\rho}^{\!~b}\omega_{\!\mu}^{\!~c}\!\Big)
\nonumber
\\
 +2\gamma \mathcal{L}_{LCS}(\omega)
\!+\!\frac{\gamma}{2\ell^{2}} e_{\mu}^{~c} D_{\nu} e_{\rho c}
\!+\!\frac{\alpha}{2\tilde{\ell}^{2}} \tilde{e}_{\mu}^{~c} \tilde{D}_{\nu} \tilde{e}_{\rho c} 
\Big\},~~~~~~~~~ \nonumber
\end{eqnarray}
where $I_{GZDG}$ is the ``general zwei-dreibein gravity" action \cite{W.Merbis,5E.A.Bergshoeff},
\begin{eqnarray}
I_{GZDG}=I_{ZDG}(\beta_{1}=0) +\frac{M_{p}}{\mu} \int d^{3}x~ L_{LCS}(\omega_{2}), 
\end{eqnarray}
with~  $\mu\!=\!\frac{\hat{m}\alpha_{2}}{\beta_{2}\sqrt{\frac{\sigma}{\alpha_{1}}}-\sqrt{\alpha_{2}}},$ 
and ``zwei-dreibein gravity" $I_{ZDG}$ has the following form \cite{3E.A.Bergshoeff,4E.A.Bergshoeff}:
\begin{eqnarray}\label{ZDG}
I_{ZDG}\!=\! -\frac{1}{2} M_{p} \!\!\int\!\! d^{3}x
\epsilon^{\mu\nu\rho} \Big\{ \sigma e_{1\mu}^{~~c} D_{\nu}
\omega_{1\rho c} \!+\! e_{2\mu}^{~~c} D_{\nu} \omega_{2\rho c}~~~  \nonumber
\\  
+\frac{1}{3} \alpha_{1} \hat{m}^{2} \epsilon_{abc} e_{1\mu}^{~~a}
e_{1\nu}^{~~b} e_{1\rho}^{~~c} \!+\!\frac{1}{3} \alpha_{2} \hat{m}^{2}
\epsilon_{abc} e_{2\mu}^{~~a} e_{2\nu}^{~~b} e_{2\rho}^{~~c}  \nonumber
\\ 
-\beta_{1} \hat{m}^{2} \epsilon_{abc} e_{1\mu}^{~~a} e_{1\nu}^{~~b}
e_{2\rho}^{~~c} \!-\!\beta_{2} \hat{m}^{2} \epsilon_{abc} e_{1\mu}^{~~a}
e_{2\nu}^{~~b} e_{2\rho}^{~~c} \Big\},~
\end{eqnarray}
where $M_{p}$ is the Planck mass, $\sigma=\pm1$ is a sign parameter, $\alpha_{1}$ and $\alpha_{2}$
are cosmological parameters, $\beta_{1}$ and $\beta_{2}$ are
coupling constants, and $e_{I\mu}^{~~a}$ and $\omega_{I\mu}^{~~a}$
($I=1,2$) are pairs of dreibein and spin connection one-forms,
respectively.
Indeed, the GZDG action is obtained as a part of our group theoretical model
(\ref{First model}), using the Chern-Simons formulation, where adding the extra terms to the GZDG action changes the GZDG massive theory to our massless model.
By varying each of the actions (\ref{First model}) or (\ref{second model}), one obtains same equations of motion as follows:
\begin{eqnarray}\label{E.O.M}
T_{\nu\rho}^{~~a}=\tilde{T}_{\nu\rho}^{~~a}=R_{\nu\rho}^{~~a} =\tilde{R}_{\nu\rho}^{~~a}=0,
\end{eqnarray}
where $T_{\nu\rho}^{~~a}, \tilde{T}_{\nu\rho}^{~~a}, R_{\nu\rho}^{~~a}$ and $\tilde{R}_{\nu\rho}^{~~a}$ are defined in (\ref{Torsion-Curvatures}).
In comparison with the general relativity, the equations of motion (\ref{E.O.M}) have some additional terms with respect to that of the general relativity, such that if one turns off the fields $\tilde{e}_{\mu}^{~a}$ and $\tilde{\omega}_{\mu}^{~a}$ in (\ref{E.O.M}), they reduce to the equations of motion of the general relativity.

%%%%%%%%%%%%%%%%%%%%%%%%%%%%%%%%%%%
\subsection {A. Anti de Sitter solution}

We obtain the following solution for the equations of motion (\ref{E.O.M}), which contains two anti de Sitter metrics:
\begin{eqnarray}\label{AdS solution-1}
ds^{2}=df^{2}=-N^2(r) dt^2+ \frac{1}{N^2(r)} dr^2 +r^2 d\varphi ^2, \nonumber
\\
N^2(r) = 1+\frac{ r^{2}}{\ell^{2}},~~~~~~~ \tilde{\ell}=\ell,~~~~~~~~~~~~~~
\end{eqnarray}
where
\begin{eqnarray}
ds^{2}=g_{\mu\nu}dx^{\mu}dx^{\nu},~~~~~~~  g_{\mu\nu}=e_{\mu}^{~a} e_{\nu}^{~b} \eta_{ab}, \nonumber
\\
df^{2}=f_{\mu\nu}dx^{\mu}dx^{\nu},~~~~~~~ f_{\mu\nu}=\tilde{e}_{\mu}^{~a} \tilde{e}_{\nu}^{~b} \eta_{ab},
\end{eqnarray}
and spin connections $\omega^{a}(r)$ and $\tilde{\omega}^{a}(r)$:
\begin{eqnarray} \label{AdS solution-2}
\omega^{0}\!=\! (1\!+\!b_{1}) N(r) \frac{dt}{\ell} \!+\!\frac{b_{1}}{b_{2}} (1\!+\!b_{1}) \frac{dr}{\ell}
\!+\! b_{1} N(r) d\varphi,~~~ \nonumber 
\\
\omega^{1}\!=\! b_{2} (\frac{dt}{\ell}\!+\! d\varphi) \!+\!\frac{(1\!+\!b_{1})}{N(r)} \frac{dr}{\ell},~~~~~~~~~~~~~~~~~~~~~~~~~~ \nonumber
\\
\omega^{2}\!=\! b_{1} \frac{r}{\ell} \frac{dt}{\ell} \!+\! \frac{b_{1}}{b_{2}}  \frac{(1\!+\! b_{1})}{N(r)} \frac{r}{\ell}\frac{dr}{\ell} 
\!+\!(1\!+\! b_{1}) \frac{r}{\ell} d\varphi,~~~~~~~~~ \nonumber
\\
\tilde{\omega}^{0}\!=\!- (2\!+\! b_{1}) N(r) \frac{dt}{\ell} \!-\!(1\!+\! b_{1})\Big(\frac{b_{1}}{b_{2}} \frac{dr}{\ell} \!+\! N(r) d\varphi\Big)\!, 
\nonumber
\\
\tilde{\omega}^{1}\!=\! - b_{2} (\frac{dt}{\ell}\!+\! d\varphi) \!-\!\frac{(2\!+\! b_{1})}{N(r)} \frac{dr}{\ell},~~~~~~~~~~~~~~~~~~~~~~~  \nonumber
\\
\tilde{\omega}^{2}\!=\! - (1\!+\! b_{1}) \frac{r}{\ell} \Big(\frac{dt}{\ell} \!+\! \frac{b_{1}}{b_{2}N(r)} \frac{dr}{\ell} \Big)
\!- \!(2\!+\! b_{1}) \frac{r}{\ell} d\varphi,~~
\end{eqnarray}
where $\{x^{0}, x^{1}, x^{2}\} = \{t, r,\varphi\}$ are the coordinates of the spacetime, and $b_{1}$ and $b_{2}$ are arbitrary constants.
We define our gravity as: $e^{a}\!=\!\tilde{e}^{a}\!\equiv\!\bar{e}^{a}$ (or equivalently $g_{\mu\nu}\!=\!f_{\mu\nu}\!\equiv\!\bar{g}_{\mu\nu}$), where $\bar{e}^{a}$ is the dreibein related to the AdS spacetime metric $\bar{g}_{\mu\nu}$.
Now, we write the quasilocal stress tensor \cite{Brown and York,Balasubramanian} defined locally on the boundary of the AdS spacetime metric $\bar{g}_{\mu\nu}$ as follows: 
\begin{eqnarray} \label{stress tensor}
T^{ij}\!=\! \frac{2}{\sqrt{-\bar{\gamma}}} (\frac{\delta I}{\delta
e_{\ell}^{~d}} \frac{\delta e_{\ell}^{~d}}{\delta \bar{\gamma}_{ij}} \!
+\!\frac{\delta I}{\delta
\tilde{e}_{\ell}^{~d}} \frac{\delta \tilde{e}_{\ell}^{~d}}{\delta \bar{\gamma}_{ij}}),
\end{eqnarray}
where $\bar{\gamma}_{ij}\!~(\! i,j\!=\!0,2\!)$ is the boundary metric of the spacetime metric $\bar{g}_{\mu\nu}$, and $\bar{\gamma}\!=\! det(\!\bar{\gamma}_{ij}\!)\!$.
Varying the action (\ref{First model}) produces a bulk term,
which is zero using the equations of motion, plus a boundary term
which can be removed by adding the following boundary term to the action (\ref{First model}):
\begin{eqnarray}
I_{B}\!=\!\frac{1}{2\pi} \!\!\int_{\!\partial\! M_{r}} \!\!\!\!\!\! d^{2}\!x \Big\{ \epsilon^{ij}\! \Big( \ell^{-1}(2\beta\!-\!\gamma) e_{i}^{~c} \omega_{jc} 
\!+\!\tilde{\ell}^{-1}(\alpha\!-\!\beta) \tilde{e}_{i}^{~c} \tilde{\omega}_{jc}
\nonumber
\\
+\beta \tilde{\omega}_{i}^{~c} (\omega_{jc}+\ell^{-1} e_{jc}) \Big)
+2\pi{\cal{L}}_{ct} \Big\},~~~~~~~~~~~~~~~~~~~~~~~
\end{eqnarray}
where we add the following counterterm
\begin{eqnarray} \nonumber
{\cal{L}}_{ct}\!=\!\frac{1}{2\pi \ell^{2}}\Big(\gamma b_{1}\!-\!\beta(b_{1}\!-\!1)   
+\frac{\ell}{\tilde{\ell}} (\alpha\!-\!\beta)(1+b_{1}) \Big) \sqrt{-\bar{\gamma}},
\end{eqnarray}
to obtain a finite stress tensor $T^{ij}$ at infinity $(r \rightarrow \infty)$
\cite{Balasubramanian}. Then, the regularized quasilocal stress tensors is:
\begin{eqnarray} \label{Stress tensor-1} 
T_{reg}^{ij}\!=\! \frac{1}{2\pi\ell \sqrt{\!-\bar{\gamma}}} ~\epsilon^{ik}\bar{\gamma}^{j\ell} \!\Big(\! (2\beta\!\!-\!\gamma) \omega_{k}^{\!~c} e_{\ell c} 
+\frac{\ell}{\tilde{\ell}} (\!\alpha\!-\!\!\beta\!) \tilde{\omega}_{\!k}^{\!~c} 
\tilde{e}_{\ell c}\!\Big)
\nonumber
\\
\!+\beta\tilde{\omega}_{k}^{\!~c} e_{\ell c} 
\!+\!\frac{1}{\!2\pi\ell^{2}\!}\!\Big(\!\gamma b_{1}\!\!-\!\beta(b_{1}\!\!-\!1)\!+\!\frac{\ell}{\tilde{\ell}} (\alpha\!-\!\beta)(1\!+\!b_{1}\!)\!\Big) \bar{\gamma}^{ij}\!\!.~~
\end{eqnarray}
The mass of the AdS solution (\ref{AdS solution-1})-(\ref{AdS solution-2}) is defined \cite{Balasubramanian} as:
\begin{eqnarray} \label{mass def.}
m\equiv\int_{0}^{2\pi} \!\!d\varphi ~r N(r) u^{i} u^{j} T^{reg}_{ij},
\end{eqnarray}
where $u^{i}=\frac{1}{N(r)}~\delta^{i,0}$ is the timelike unit
normal to spacelike surface at the boundary of the AdS spacetime.
The mass equals the value of the integral when it is computed at $(r\!\rightarrow\! \infty)$, 
\begin{eqnarray} \label{AdS mass}
m\!=\!\!\frac{1}{2\ell}\Big(\!(\gamma \!-\! 2\beta) b_{1}\!+\!\alpha (\!1\!+ b_{1}\!)\Big)\!.
\end{eqnarray}
By substituting $\alpha=\gamma=0$ in (\ref{AdS mass}), which is obtained for the model (\ref{First model}), one obtains the mass of the AdS solution for the model (\ref{second model}).

%%%%%%%%%%%%%%%%%%%%%%%%%%%%%%%%%%%%%
\section{III. $AdS_{3}$/$CFT_{2}$ correspondence}

In this section, we show that (2+1)-dimensional bi-gravity model (\ref{First model}) has a two dimensional dual conformal field theory at the boundary of the AdS spacetime. 
Using the following redefinitions of the generators:
\begin{eqnarray}
W_{a}^{+}\!=\!\frac{1}{2}(J_{a}\!+\!\ell P_{a}\!-\!\tilde{J}_{a}\!+\!\tilde{\ell} \tilde{P}_{a}),~~~
W_{a}^{-}\!=\!\frac{1}{2}(\tilde{J}_{a}\!-\!\tilde{\ell} \tilde{P}_{a}),~\nonumber
\\
\widetilde{W}_{a}\!=\!\frac{1}{2}(2\tilde{J}_{a}\!-\!J_{a}\!-\!\ell P_{a}),~~~~~~~~~
\overline{W}_{a}\!=\!\frac{1}{2}(J_{a}\!-\!\ell P_{a}),~~
\end{eqnarray}
the mixed AdS-AdS Lie algebra (\ref{mixed AdS-AdS Lie algebra}) can be rewritten as follows:
\begin{eqnarray} \label{direct sum}
[W_{a}^{\pm},W_{b}^{\pm}] \!=\! \epsilon_{abc}W^{\pm c}\!,~~~~~
[\overline{W}_{a},\overline{W}_{b}] \!=\! \epsilon_{abc}\overline{W}^{c}\!, \nonumber
\\ ~
[\widetilde{W}_{a},\widetilde{W}_{b}] \!=\! \epsilon_{abc}\widetilde{W}^{c}\!,~~~~~~~~~
[W_{a}^{+},W_{b}^{-}]\!=\! 0,~~~~~~ \nonumber
\\ ~
[\overline{W}_{a},\widetilde{W}_{b}] \!=\!
[W_{a}^{\pm},\overline{W}_{b}] \!=\!
[W_{a}^{\pm},\widetilde{W}_{b}] \!=\! 0,~~~~~~~~~~
\end{eqnarray}
which is clearly a direct sum of four $sl(2,\mathbb{R})$ algebras (i.e. the mixed AdS-AdS Lie algebra is isomorphic to $sl(2,\mathbb{R})\oplus sl(2,\mathbb{R})\oplus sl(2,\mathbb{R})\oplus sl(2,\mathbb{R})$).
In terms of the new generators, we rewrite the Lie algebra valued gauge field (\ref{Murer-Cartan one-form}) as follows:
\begin{eqnarray} \nonumber
h_{\mu}\!=\! h_{\mu}^{~B}W_{B}\!=\! h_{\mu}^{+ a}~ W_{a}^{+} \!+\! h_{\mu}^{- a}~ W_{a}^{-} \!+\! \tilde{h}_{\mu}^{~a}~ \widetilde{W}_{a}\!+\! \bar{h}_{\mu}^{~a}~ \overline{W}_{a},
\end{eqnarray}
where we have:
\begin{eqnarray} \label{new gauge field}
h_{\mu}^{\pm a}=\omega_{\mu}^{~a}+\frac{1}{\ell} e_{\mu}^{~a}+\tilde{\omega}_{\mu}^{~a}\pm\frac{1}{\tilde{\ell}} \tilde{e}_{\mu}^{~a}, ~~~~~~~~~~~~~ \nonumber
\\
\tilde{h}_{\mu}^{~a}=\tilde{\omega}_{\mu}^{~a}+\frac{1}{\tilde{\ell}} \tilde{e}_{\mu}^{~a},~~~~~~~~
\bar{h}_{\mu}^{~a}=\omega_{\mu}^{~a}-\frac{1}{\ell} e_{\mu}^{~a}.
\end{eqnarray}
Then, using $x^{\pm}=\varphi\pm \frac{t}{\ell}$, and a suitable $2\times 2$ matrix basis for the $sl(2,\mathbb{R})$ algebras, we rewrite the anti de Sitter solution (\ref{AdS solution-1})-(\ref{AdS solution-2}) as follows:
\begin{eqnarray} \label{matrix form of fields}
h^{\pm}\!\!=\!\frac{1}{2}\! \left( \begin{tabular}{cc}
$\pm\frac{1}{\ell N(r)} dr$ & $-\Big(N(r)\mp\frac{r}{\ell}\Big) dx^{\mp}$ \\
$\Big(N(r)\pm\frac{r}{\ell}\Big) dx^{\mp}$ & $\mp\frac{1}{\ell N(r)} dr$ \\
\end{tabular} \right)\!,~~~~~~~  \nonumber
\\
\bar{h}\!=\! \frac{1}{2}\! \tiny{\left( \begin{tabular}{cc}
$\frac{b_{1}}{\ell N(r)} dr\!+\!b_{2}dx^{+}$ & $\xi^{+}\!(r) \Big( b_{1}dx^{+}\!+\!\chi(r)dr\Big)$ \\
$-\xi^{-}\!(r) \Big( b_{1}dx^{+}\!+\!\chi(r)dr\Big)$ & $-\frac{b_{1}}{\ell N(r)} dr\!-\! b_{2}dx^{+}$ \\
\end{tabular} \right)\tiny}\!,~~~~~~~~~~  \nonumber
\\
\tilde{h}\!=\! \frac{1}{2}\! \tiny{\left( \begin{tabular}{cc}
$-\frac{(1\!+\!b_{1})}{\ell N(r)} dr\!-\!b_{2}dx^{+}$ & $-\xi^{+}\!(r) \Big(\! (1+b_{1})dx^{+}\!+\!\chi(r)dr\!\Big)$ \\
$\xi^{-}\!(r) \Big(\! (1\!+\!b_{1})dx^{+}\!+\!\chi(r)dr\!\Big)$ & $\frac{(1\!+\!b_{1})}{\ell N(r)} dr\!+\!b_{2}dx^{+}$ \\
\end{tabular} \right)\tiny}\!,\nonumber
\\
\xi^{\pm}(r)\equiv N(r)\pm\frac{r}{\ell},~~~~~~~
\chi(r)\equiv \frac{ b_{1}(1+b_{1})}{b_{2}\ell N(r)},~~~~~~~
\end{eqnarray}
from which we see that $h_{+}^{+}=h_{-}^{-}=\bar{h}_{-}=\tilde{h}_{-}=0,$ and then we have:
\begin{eqnarray}\label{boundary condition} 
h_{0}^{+}\!=\!-\frac{1}{\ell} h_{2}^{+},~~~
h_{0}^{-}\!=\!\frac{1}{\ell} h_{2}^{-},~~~
\bar{h}_{0}\!=\!\frac{1}{\ell} \bar{h}_{2},~~~
\tilde{h}_{0}\!=\!\frac{1}{\ell} \tilde{h}_{2}.~
\end{eqnarray}
Using (\ref{direct sum})-(\ref{matrix form of fields}), the bi-gravity action (\ref{First model}) can be rewritten as a sum of four $sl(2,\mathbb{R})$ Chern-Simons theories as follows:
\begin{eqnarray}\label{sum of four C-S} 
I \!=\! K^{+} I_{cs}(h_{\mu}^{+}) \!+\! K^{-} I_{cs}(h_{\mu}^{-}) \!+\! \bar{K}  I_{cs}(\bar{h}_{\mu})\!+\! \tilde{K}  I_{cs}(\tilde{h}_{\mu}),~
\end{eqnarray}
where $K^{\pm}=\frac{\beta}{2},~\bar{K}=\gamma-\beta$ and $\tilde{K}=\alpha-\beta$ are levels of the Chern-Simons actions which three of them are independent, only \footnote{The non-degeneracy conditions of the bilinear quadratic form (\ref{adinvariant metric}) on the mixed AdS-AdS Lie algebra (i.e. $\beta\neq 0$, $\beta\neq \alpha$ and $\beta\neq \gamma$) imply that all of four levels of the Chern-Simons actions (i.e. $K^{\pm},~\bar{K},~\tilde{K}$) are non-zero. The mixed AdS-AdS Lie algebra (\ref{mixed AdS-AdS Lie algebra}) and its bilinear quadratic form (\ref{adinvariant metric}) requires that the most general interacting theory must have three independent parameters, and then in the most general theory, three of the levels are independent. But if we require that one of the parameters $\alpha$ or $\gamma$ vanishes, then we will have two independent levels, while by requiring that both of the parameters $\alpha$ and $\gamma$ vanish, we will only have one independent level in the theory.}. Each of the Chern-Simons actions $I_{cs}(h_{\mu}^{\pm}),~I_{cs}(h_{\mu})$ and $I_{cs}(\tilde{h}_{\mu})$ up to a surface term can be written as:
\begin{eqnarray}\label{C-S action}
I_{cs}(h_{\mu})\!=\!\frac{1}{4\pi} \!\!\int\!\! d^{3}x \Big[
h_{2}^{~a} \partial_{0} h_{1a} - h_{1}^{~a}
\partial_{0} h_{2a} + 2 h_{0}^{~c} F_{12a} \Big], \nonumber
\\
F_{12a}= \partial_{1} h_{2a} - \partial_{2}
h_{1a} + \epsilon_{abc} h_{1}^{~b} h_{2}^{~c}.~~~~~~~~~~~~~~~~~~~~
\end{eqnarray}
Variation of each of the Chern-Simons actions $I_{cs}(h_{\mu}^{\pm}),~I_{cs}(\bar{h}_{\mu})$ and $I_{cs}(\tilde{h}_{\mu})$ do not vanishes at the boundary, even when the equations of motion hold, because of $\int d^{3}x ~Tr(h_{0} \partial_{1} h_{2})$ term. To obtain an action with zero variation ($\delta I=0$), using (\ref{boundary condition}) we add a surface term to the action (\ref{sum of four C-S}) as follows:
\begin{eqnarray} \label{model plus a surface term} 
I = K^{+} I_{cs}(h_{\mu}^{+}) + K^{-} I_{cs}(h_{\mu}^{-}) + \bar{K} I_{cs}(\bar{h}_{\mu})+ \tilde{K} I_{cs}(\tilde{h}_{\mu}) 
\nonumber
\\
\! +\frac{1}{4\pi\ell}\! \!\int_{\Sigma}\!\! \! dt d\varphi T\!r\!\Big[\! K^{+}\!(h_{2}^{\!+}\!)^2 \! \!-\! K^{-}(h_{2}^{\!-}\!)^{\! 2} \! \!-\!\bar{K} (\bar{h}_{2}\!)^2 
\! -\!\tilde{K}\! (\tilde{h}_{2}\!)^2\!\Big]\! ,~
\end{eqnarray}
where $\Sigma$ is a two dimensional boundary. By variation of the action (\ref{model plus a surface term}) with respect to the gauge fields $h_{0}^{\pm}$, $\bar{h}_{0}$ and $\tilde{h}_{0}$ as Lagrange multipliers, one obtains four constraints, $F_{12}^{\pm}=\bar{F}_{12}=\tilde{F}_{12}= 0,$ with the following solution: \begin{eqnarray}\label{solution of constraints} 
h_{\mu}^{+} = G_{1}^{-1} \partial_{\mu} G_{1},~~~~~~~ 
h_{\mu}^{-} = G_{2}^{-1} \partial_{\mu} G_{2},~  \nonumber
\\
\bar{h}_{\mu}= G_{3}^{-1} \partial_{\mu} G_{3},~~~~~~~
\tilde{h}_{\mu}= G_{4}^{-1} \partial_{\mu} G_{4},~
\end{eqnarray}
where $G_{i}(t,r,\varphi)$,~$i=1...4$~ are as follows:
\begin{eqnarray}
G_{1}\!=\! g_{1}(t,\varphi) \small{\left( \begin{tabular}{cc}
$\!U^{+}\!(r)$ & $0$ \\
$0$ & $\frac{1}{U^{+}\!(r)}\!$ \\
\end{tabular} \right)\small}\!,~~~~  U^{\pm}(r)\!\equiv\! \sqrt{\frac{r}{\ell}\pm N(r)},~ \nonumber
\\
G_{2}= g_{2}(t,\varphi) \small{\left( \begin{tabular}{cc}
$\frac{1}{U^{+}\!(r)}$ & $0$ \\
$0$ & $U^{+}\!(r)$ \\
\end{tabular} \right)\small}\!,~~~~~~~~~~~~~~~~~~~~~~~~~~~~~~~~~ \nonumber
\\
G_{3}= g_{3}(t,\varphi) \small{\left( \begin{tabular}{cc}
$b_{1}\Big(U^{-}\Big)^{\!b^{+}}$ & $d_{1}\Big(U^{-}\Big)^{\!b^{-}}$ \\
$d_{1}\Big(U^{-}\Big)^{\!-b^{-}}$ & $b_{1}\Big(U^{-}\Big)^{\!-b^{+}}$ \\
\end{tabular} \right)\small}\!,~~~~~~~~~~~~~~~ 
 \nonumber
\\
G_{4}\!=\! g_{4}(t,\varphi) \small{\left( \begin{tabular}{cc}
$(b_{1}\!+\!1)\Big(\!U^{-}\!\Big)^{\!e^{+}}$ & $d_{2}\Big(\!U^{-}\!\Big)^{\!e^{-}}$ \\
$d_{2}\Big(\!U^{-}\!\Big)^{\!-e^{-}}$ & $(b_{1}\!+\!1)\Big(\!U^{-}\!\Big)^{\!-e^{+}}$ \\
\end{tabular} \right)\small}\!,~~~~~~~
\end{eqnarray}
where $g_{i}(t,\varphi)$,~$i=1...4$~ are arbitrary elements of the Lie group $SL(2,\mathbb{R})$ and we have:
\begin{eqnarray} \label{cte}
d_{1}\equiv b_{2}+\sqrt{(b_{2})^2-(b_{1})^2},~~~~~~~ \nonumber  \\
d_{2}\equiv b_{2}-\sqrt{(b_{2})^2-(b_{1}+1)^2},~~ \nonumber  \\
b^{\pm}\equiv \frac{b_{1}+1}{b_{2}} \sqrt{(b_{2})^2-(b_{1})^2}\pm 1, \nonumber \\
e^{\pm}\equiv \frac{b_{1}}{b_{2}} \sqrt{(b_{2})^2-(b_{1}+1)^2}\pm 1.
\end{eqnarray}
Then, using (\ref{solution of constraints})-(\ref{cte}) the surface term in (\ref{model plus a surface term}) can be rewritten in the following form:
\[
\frac{1}{4\pi \ell} \int_{\Sigma} dt d\varphi 
Tr\Big( K^{+} (g'_{1})^{2} -K^{-} (g'_{2})^{2} -\bar{K} (g'_{3})^{2} 
-\tilde{K} (g'_{4})^{2} \Big),~
\]
where $g'_{i}\equiv g_{i}^{-1} \partial_{2} g_{i}~(i=1...4)$, and the action (\ref{model plus a surface term}) is given by
\begin{eqnarray}
I= K^{+} S_{WZW}^{L}[g_{1}] + K^{-} S_{WZW}^{R}[g_{2}] 
\nonumber
\\
+\bar{K}~ S_{WZW}^{R}[g_{3}] +\tilde{K}~ S_{WZW}^{R}[g_{4}],
\end{eqnarray}
where $S_{WZW}^{L}[g_{1}]$ and $S_{WZW}^{R}[g_{m}]$, $(m=2,3,4)$ are chiral Wess-Zumino-Witten actions over $SL(2,\mathbb{R})$ group which
describe a left-moving group element $g_{1}(x^{-})$ together with three right-moving group elements $g_{2}(x^{+})$, $g_{3}(x^{+})$ and $g_{4}(x^{+})$,
respectively, and have the following forms:
\begin{eqnarray}
S_{W\!ZW}^{L}[g_{1}\!]\!=\! \frac{1}{4\pi} \!\!\int_{\Sigma}\!\!\! dt d\varphi
T\!r\Big[ g'_{1}\dot{g}_{1} \!+\!\frac{1}{\ell} (g'_{1})^2 \Big] \!+\!
\Gamma[g_{1}\!],~~~
\nonumber
\\
S_{W\!ZW}^{R}[g_{m}\!]\!=\! \frac{1}{4\pi} \!\!\int_{\Sigma}\!\!\! dt d\varphi
T\!r\Big[ g'_{m}\dot{g}_{m} \!-\! \frac{1}{\ell} (g'_{m})^2 \Big] \!+\!
\Gamma[g_{m}\!], 
\end{eqnarray}
where $\dot{g}_{i}\equiv g_{i}^{-1}\partial_{0} g_{i}~(i=1...4)$, and $\Gamma[g]$'s are the usual Wess-Zumino terms of the Wess-Zumino-Witten actions as follows:
\begin{eqnarray}  
\Gamma[g_{i}]= -\frac{1}{12\pi} \int_{M} Tr\Big(G_{i}^{-1} dG_{i}\Big)^3,~~~~~ i=1...4.
\end{eqnarray}
Thus far, we have shown that the bi-gravity model (\ref{First model}) has a dual CFT at the boundary of the AdS spacetime, which is the sum of four $SL(2,\mathbb{R})$ WZW actions.
It is obvious (by substituting $\alpha=\gamma=0$) that the other model (\ref{second model}) has also $AdS_{3}/CFT_{2}$ correspondence at the boundary of the AdS spacetime.

%%%%%%%%%%%%%%%%%%%%%%%%%%%%%%%%%%%%%
\subsection{A. Central charge of the conformal field theory}

Now, we calculate the central charge ``$c$'' of the discussed two dimensional conformal field theory using the following relation: \cite{Francesco,Henningson}
\begin{eqnarray} \label{central charge relation}
Tr(T^{ij})=-\frac{c}{24\pi}\mathcal{R},
\end{eqnarray}
where $T^{ij}$ and $\mathcal{R}$ are the regularized quasilocal stress tensor (\ref{Stress tensor-1}) and scalar curvature of the boundary surface, respectively. $\mathcal{R}$ can be calculated using the following identity: \cite{Hasanpour}
\begin{eqnarray}
G_{ij} n^{i} n^{j}=-\frac{1}{2} (\mathcal{R}+\theta_{ij} \theta^{ij}-\theta^{2}),
\end{eqnarray}
where $G_{ij}$ is the Einstein tensor, $n^{i}=\frac{1}{\sqrt{\bar{g}_{rr}}} \delta^{i,r}$ is a unit outward pointing normal vector to the boundary, and $\theta_{ij}=-\frac{1}{2\sqrt{\bar{g}_{rr}}} \partial_{r} \bar{\gamma}_{ij}$ is the extrinsic curvature of the boundary metric $\bar{\gamma}_{ij}$, which trace is $\theta=\bar{\gamma}^{ij}\theta_{ij}$.
Using the Fefferman-Graham expansion of the boundary metric \cite{Fefferman} 
\begin{eqnarray}\label{Fefferman-Graham}
\bar{\gamma}_{ij}=r^{2} \bar{\gamma}^{(0)}_{ij}+\bar{\gamma}^{(2)}_{ij} + O(\frac{1}{r^{2}}), ~~~
\bar{\gamma}^{(0)}=diag(\ell^{-2},1),
\end{eqnarray}
one obtains the scalar curvature of the boundary surface at infiniry $(r\rightarrow\infty)$ as follows:
\begin{eqnarray} \label{boundary curvature}
\mathcal{R}=\frac{2}{\ell^2 r^{2}} \bar{\gamma}^{(0)^{ij}} \bar{\gamma}^{(2)}_{ij} +\cdot\cdot\cdot~.
\end{eqnarray}
By use of the Fefferman-Graham expansion (\ref{Fefferman-Graham}) we obtain the following relation:
\begin{eqnarray} 
\frac{1}{\sqrt{-\bar{\gamma}}}=\frac{\ell}{r^{2}} ~\Big( 1
-\frac{1}{2r^{2}} \bar{\gamma}^{(0)^{ij}}
\bar{\gamma}^{(2)}_{ij}+\cdot\cdot\cdot \Big),
\end{eqnarray}
using which together with (\ref{Stress tensor-1}), trace of the regularized quasilocal stress tensor $T_{reg}^{ij}$ at boundary $(r\rightarrow\infty)$ of the AdS spacetimes is given by
\begin{eqnarray}\label{trace of stress tensor} 
Tr(T_{reg}^{ij}) 
\!=\!\frac{(\gamma \!-\! 2\beta) b_{1}\!+\!\alpha (1\!+\! b_{1})}{2\pi \ell^2 r^{2}} ~\bar{\gamma}^{(0)^{ij}} \bar{\gamma}^{(2)}_{ij} 
\!+\!\cdot\cdot\cdot~.~~
\end{eqnarray}
By inserting (\ref{boundary curvature}) and (\ref{trace of stress tensor}) in (\ref{central charge relation}), one obtains the central charge of the boundary CFT as follows:
\begin{eqnarray}
c\!=\!-6 [(\gamma \!-\! 2\beta) b_{1}\!+\!\alpha (1\!+\! b_{1}) ],
\end{eqnarray}
which is proportional to the mass ($m$) of the AdS solution (\ref{AdS mass}):
\begin{eqnarray}\label{central charge-mass}
c=-12\ell m,
\end{eqnarray}
where a negative mass of the AdS solution leads to a positive central charge.
The same calculations show that the relation (\ref{central charge-mass}) is also valid for the ZDG model (\ref{ZDG}). Indeed, the mass of the AdS solution $\tilde{e}^{a}\!=\!\hat{\gamma}e^{a}$ and $\tilde{\omega}^{a}\!=\!\omega^{a}$ of the ZDG model is
$m_{\scriptscriptstyle{Z\!DG}}\!=\!- M_{p} \pi(\sigma+\hat{\gamma})$, and the central charge of the boundary CFT for the ZDG model is $c_{\scriptscriptstyle{Z\!DG}}\!=\!12\pi\ell M_{p}(\sigma+\hat{\gamma})$ \cite{3E.A.Bergshoeff,W.Merbis}.

%%%%%%%%%%%%%%%%%%%%%%%%%%%%%%%%%%%%%
\section{IV. Cosmological implications}

In this section, we study homogeneous and isotropic cosmology of our massless bi-gravity theory (\ref{First model}) by use of the Friedmann-Robertson-Walker (FRW) Ansatz for both metrics:
\begin{eqnarray}\label{FRW-1}
ds^{2}=-N^{2}(t) ~dt^{2}+a^{2}(t) \Big(\frac{dr^{2}}{1-k r^2}+r^{2}d\varphi^{2}\Big),
\end{eqnarray}
and
\begin{eqnarray}\label{FRW-2}
df^{2}=-X^{2}(t) ~dt^{2}+Y^{2}(t) \Big(\frac{dr^{2}}{1-k r^2}+r^{2}d\varphi^{2}\Big),
\end{eqnarray}
where $a(t)$ and $Y(t)$ are the spatial scale factors of the metrics $g_{\mu\nu}$ and $f_{\mu\nu}$, respectively, and $N(t)$ and $X(t)$ are their lapse functions.
The constant $k$ in both metrics is the spatial curvature, whose negative, vanishing and positive values ($k=-1,0,1$) correspond to open, flat and closed universes, respectively.

We use the FRW metrics (\ref{FRW-1})-(\ref{FRW-2}) to solve the equations of motion (\ref{E.O.M}), and obtain the following (Friedmann) equations:
\begin{eqnarray}\label{Friedmann-1}
\Big(\frac{\dot{a}}{a}\Big)^{2}=\tilde{\Lambda}(1-\frac{k}{\Lambda a^{2}}) X^{2},
\end{eqnarray}
\begin{eqnarray}\label{Friedmann-2}
\frac{\ddot{a}}{a}=\frac{\dot{a}}{a}\frac{\dot{X}}{X}+\tilde{\Lambda} X^{2},
\end{eqnarray}
together with a relation between two scale factors as:
\begin{eqnarray}
Y(t)=\sqrt{\frac{\Lambda}{\tilde{\Lambda}}}~a(t),
\end{eqnarray}
and the following relations for spin connection fields:
\begin{eqnarray} \label{spin connection of FRW}
\omega^{0}\!=\! (\frac{2X}{\tilde{\ell}} - \frac{N}{\ell}) dt, ~~~~~~~~~~~~~~~~~~~~~~~~~~ 
\nonumber 
\\
\omega^{1}\!=\frac{a}{1-k r^2} \frac{dr}{\ell},~~~~~~~~~~~~~~~~~~~~~~~~~~~~ \nonumber
\\
\omega^{2}\!=  \frac{a r}{\ell} d\varphi, ~~~~~~~~~~~~~~~~~~~~~~~~~~~~~~~~~~~ \nonumber
\\
\tilde{\omega}^{0}\!=\! -\frac{2X}{\tilde{\ell}} ~dt \!-\! \sqrt{1-k r^2} ~d\varphi, ~~~~~~~~~~~~
\nonumber
\\
\tilde{\omega}^{1}\!=\!-\frac{2 a}{\sqrt{1-k r^2}} \frac{dr}{\ell}
\!+\! \frac{r\tilde{\ell}~\dot{a}}{\ell ~X}d\varphi, ~~~~~~~~~~~~\nonumber
\\
\tilde{\omega}^{2}\!=\!-\frac{\tilde{\ell}\dot{a}}{\ell X\sqrt{1-k r^2}} ~dr
\!-\! 2 \frac{a r}{\ell} ~d\varphi, ~~~~~~~
\end{eqnarray}
where $\ell^{-2}\!=\!-\Lambda$, $\tilde{\ell}^{-2}\!=\!-\tilde{\Lambda}$, and dot denotes the time derivative ($\dot{a}\equiv \frac{da}{dt},~\ddot{a}\equiv \frac{d^{2}a}{dt^{2}}$).
Using the Friedmann equations (\ref{Friedmann-1})-(\ref{Friedmann-2}), we obtain the following relation for $X(t)$ in terms of the scale factor $a(t)$:
\begin{eqnarray}\label{lapse}
X(t)=\pm \frac{\tilde{\ell}\dot{a}}{\sqrt{-k\ell^{2}- a^{2}}},~~~~~~ |a(t)|\leq \ell\sqrt{-k},
\end{eqnarray}
which implies that we have an open universe with negative spatial
curvature ($k=-1$), where the radial coordinate $r$ is defined on $0\leq r<+\infty$.
The Friedmann equations do not restrict the lapse function $N(t)$ and the scale factor $a(t)$ of the FRW metric (\ref{FRW-1}), and then $N(t)$ and $a(t)$ are arbitrary functions of the timelike coordinate $t$.
In what follows, we discuss three different possible selections of the scale factor $a(t)$, and the lapse function $N(t)$.
\\

%%%%%%%%%%%%%%%%%%%
\subsection{A. Self-accelerating solution}

Here, we have no matter in our theory (vanishing matter density $\rho=0$), and then our cosmological analysis entirely describes a late time cosmology.
By assuming that the scale factor $a(t)$ has an exponential form as follows:
\begin{eqnarray}
a(t)=a(0)~ e^{b t},
\end{eqnarray}
we obtain the following relation for the lapse function $X(t)$ using (\ref{lapse}):
\begin{eqnarray}
X(t)=\pm \frac{ a(0) \tilde{\ell} b~ e^{b t}}{\sqrt{\ell^{2}- a^{2}(0) e^{2 b t}}},
\end{eqnarray}
where $a(0)$ is the initial value of the scale factor, and $b$ is an arbitrary constant. The Hubble parameter for this solution is obtained as follows:
\begin{eqnarray}
H\equiv \frac{\dot{a}}{a}=b,
\end{eqnarray}
which is a constant. At late time, this solution is a self-accelerating solution whose deceleration parameter is as follows:
\begin{eqnarray}
q\equiv -\frac{a\ddot{a}}{\dot{a}^{2}}=-1,
\end{eqnarray}
which is obviously negative and implies that the expansion of the universe is accelerating.
\\

%%%%%%%%%%%%%%%%%%%
\subsection{B. Proportional solution}

By assuming that the lapse function $N(t)$ in (\ref{FRW-1}) is proportional to $X(t)$:
\begin{eqnarray}
N(t)=\frac{\ell}{\tilde{\ell}}~X(t),
\end{eqnarray}
two metrics $g_{\mu\nu}$ and $f_{\mu\nu}$ will be proportional as follows:
\begin{eqnarray}
ds^{2}=-\frac{\ell^{2}\dot{a}^{2}}{\ell^{2}- a^{2}} dt^{2}+a^{2} \Big(\frac{dr^{2}}{1-k r^2}+r^{2}d\varphi^{2}\Big),
\end{eqnarray}
\begin{eqnarray}
df^{2}=\frac{\tilde{\ell}^{2}}{\ell^{2}}~ds^{2}, ~~~~~~~~~~~~~~~~~~~~~~~~~~~~~~~~~~~~~~~
\end{eqnarray}
where the scale factor $a(t)$ is an arbitrary function and admits various forms such as exponential form, oscillating form and so on.
An example of such a proportional solution is a hyperbolic scale factor:
\begin{eqnarray}
a(t)=\ell~tanh(t/\ell),
\end{eqnarray}
which together with $N(t)=\frac{1}{cosh^{2}(t/\ell)},$ leads to the following proportional FRW metrics:
\begin{eqnarray}
ds^{2}\!=\! -\frac{dt^{2}}{cosh^{2}(t/\ell)} \!+\!\ell^{2}tanh^{2}(t/\ell) \Big(\!\frac{dr^{2}}{ 1\!-\! k r^2\!}\!+\! r^{2}d\varphi^{2}\!\Big),
\end{eqnarray}
\begin{eqnarray}
df^{2}=\frac{\tilde{\ell}^{2}}{\ell^{2}}~ds^{2}. ~~~~~~~~~~~~~~~~~~~~~~~~~~~~~~~~~~~~~~~~~~~~~~~~
\end{eqnarray}
This solution has a time-dependent Hubble parameter,
\begin{eqnarray}
H(t)=\frac{1}{\ell} sech(t/\ell)~ csch(t/\ell),
\end{eqnarray}
and a non-negative deceleration parameter,
\begin{eqnarray}
q(t)=2~ sinh^{2}(t/\ell),
\end{eqnarray}
which means that the expansion of the universe is decelerating.
\\

%%%%%%%%%%%%%%%%%%%
\subsection{C. Oscillating solution}

By assuming that the scale factor $a(t)$ has the following oscillating form:
\begin{eqnarray}
a(t)=\ell ~ sin(t/\ell),
\end{eqnarray}
and $N(t)=1,$ using (\ref{lapse}) one obtains the following constant value of the lapse function $X(t)$:
\begin{eqnarray}
X(t)=\pm\frac{\tilde{\ell}}{\ell}.
\end{eqnarray}
This selection of the scale factor $a(t)$ leads to the following oscillating and proportional solution:
\begin{eqnarray}
ds^{2}=- dt^{2}+\ell^{2}~sin^{2}(t/\ell) \Big(\frac{dr^{2}}{1-k r^2}+r^{2}d\varphi^{2}\Big),
\end{eqnarray}
\begin{eqnarray}
df^{2}=\frac{\tilde{\ell}^{2}}{\ell^{2}}~ds^{2}. ~~~~~~~~~~~~~~~~~~~~~~~~~~~~~~~~~~~~~~~~~
\end{eqnarray}
The Hubble parameter and the deceleration parameter of this solution are as follows:
\begin{eqnarray}
H(t)=cot(t/\ell),~~~~~~~ q(t)= tan^{2}(t/\ell),
\end{eqnarray}
where the non-negative deceleration parameter implies that the expansion of the universe is decelerating.

%%%%%%%%%%%%%%%%%%%%%%%%%%%%%%%%%%%%%
\section{V. Conclusions}

We have introduced a new Lie algebra in (2+1)-dimensional spacetime which has a pair of AdS subalgebras. By gauging the algebra, we have proposed a new Chern-Simons bi-gravity model in 2+1 spacetime dimensions which describes two interacting massless spin-2 fields \footnote{After submitting this paper, we noticed that a three dimensional theory for the colored
massless spin-two fields has been studied in \cite{Gwak}.}.
We have shown that the model has a dual CFT at the boundary of the AdS space-time such that its central charge is proportional to the mass of the AdS spacetime.
Our bi-gravity model admits some homogeneous and isotropic FRW cosmological solutions with some different scale factors, which describe both accelerating and decelerating universes.
Study of (3+1)-dimensional version of the mixed AdS-AdS Lie algebra, and resultant (3+1)-dimensional gauge invariant interacting model and its cosmological implications is also a useful task which may have interesting features.
Chern-Simons formulation of our interacting model simplifies its quantization, which may be interesting in the context of the quantum gravity.

%%%%%%%%%%%%%%%%%%%%%%%%%%%%%%%%%%%
\textbf{Acknowledgments:}
We would like to express our heartfelt gratitude to M.M. Sheikh-Jabbari, F. Darabi and F. Loran for their useful comments and discussions.

%%%%%%%%%%%%%%%%%%%%%%%%%%%%%%%%%%%%% 

%%%%%%%%%%%%%%%%%%%%%%%%%%%%%%%%%%%%
%%%%%%%%%%%%%%%%%%%%%%%%%%%%%%%%%%%%
%%%%%%%%%%%%%%%%%%%%%%%%%%%%%%%%%%%%%
%%%%%%%%%%%%%%%%%%%%%%%%%%%%%%%%%%%%%%
%\section{\label{sec:level1}First-level heading}
% sections are not used for PRL papers

%\subsection{\label{sec:level2}Second-level heading: Formatting}
% subsections are not used for PRL papers

%\input acknowledgement.tex   % input acknowledgement

\end{document}